\documentclass{snmp2003}

\input epsf.sty

\begin{document}
\newcommand{\wit}[1]{\widetilde{#1}}    
\renewcommand{\vec}[1]{\boldsymbol{#1}} 

\FirstPageHeading{Henkel}

\ShortArticleName{Dynamical space-time symmetry for ageing systems} 

\ArticleName{Dynamical space-time symmetry for ageing far from equilibrium}

\Author{Malte HENKEL}
\AuthorNameForHeading{M. Henkel}
\AuthorNameForContents{HENKEL, M.}
\ArticleNameForContents{Dynamical space-time symmetry for ageing far from
equilibrium}

\Address{Laboratoire de Physique des Mat\'eriaux CNRS UMR 7556, 
Universit\'e Henri Poincar\'e Nancy I, B.P. 239, 
F -- 54506 Vand{\oe}uvre-l\`es-Nancy, France}

\Abstract{The dynamical scaling of ageing ferromagnetic 
systems can be generalized to a local scale invariance. This yields a 
prediction for the causal two-time response function, which has been
numerically confirmed in the Glauber-Ising model quenched into the ordered
phase. For a dynamical exponent $z=2$, a new embedding of the Schr\"odinger
group into the conformal group and the resulting conditions for the validity 
of local scale invariance are discussed.}

\section{Phenomenology of ageing in simple ferromagnets}

Ageing phenomena provide a paradigmatic example of collective behaviour far
from equilibrium and have received a lot of attention in recent years 
\cite{Henkel:Cate00,Henkel:Cugl02,Henkel:Godr02}. Ageing
has been observed first in glassy systems, but for an improved conceptual
understanding, it might be more useful to study first ageing phenomena
in the simpler ferromagnetic systems, as we shall do here. We consider a
ferromagnet with a critical temperature $T_{\rm c}>0$ and prepare the system
in some initial state (which typically may be disordered). Then the system
is suddenly brought into contact (quenched) with a heat bath of temperature 
$T<T_{\rm c}$ (or $T\leq T_{\rm c}$). Keeping $T$ fixed, the system's 
evolution towards its equilibrium state at temperature $T$ is observed. For 
definiteness, consider an Ising-like system with a microscopic degree of 
freedom $\sigma_{\vec{i}}=\pm 1$ which can only take two possible values, and 
where $\vec{i}$ denotes the sites of a (hypercubic) lattice. 

\begin{figure}[h]
\centerline{\epsfxsize=2.1in\ \epsfbox{
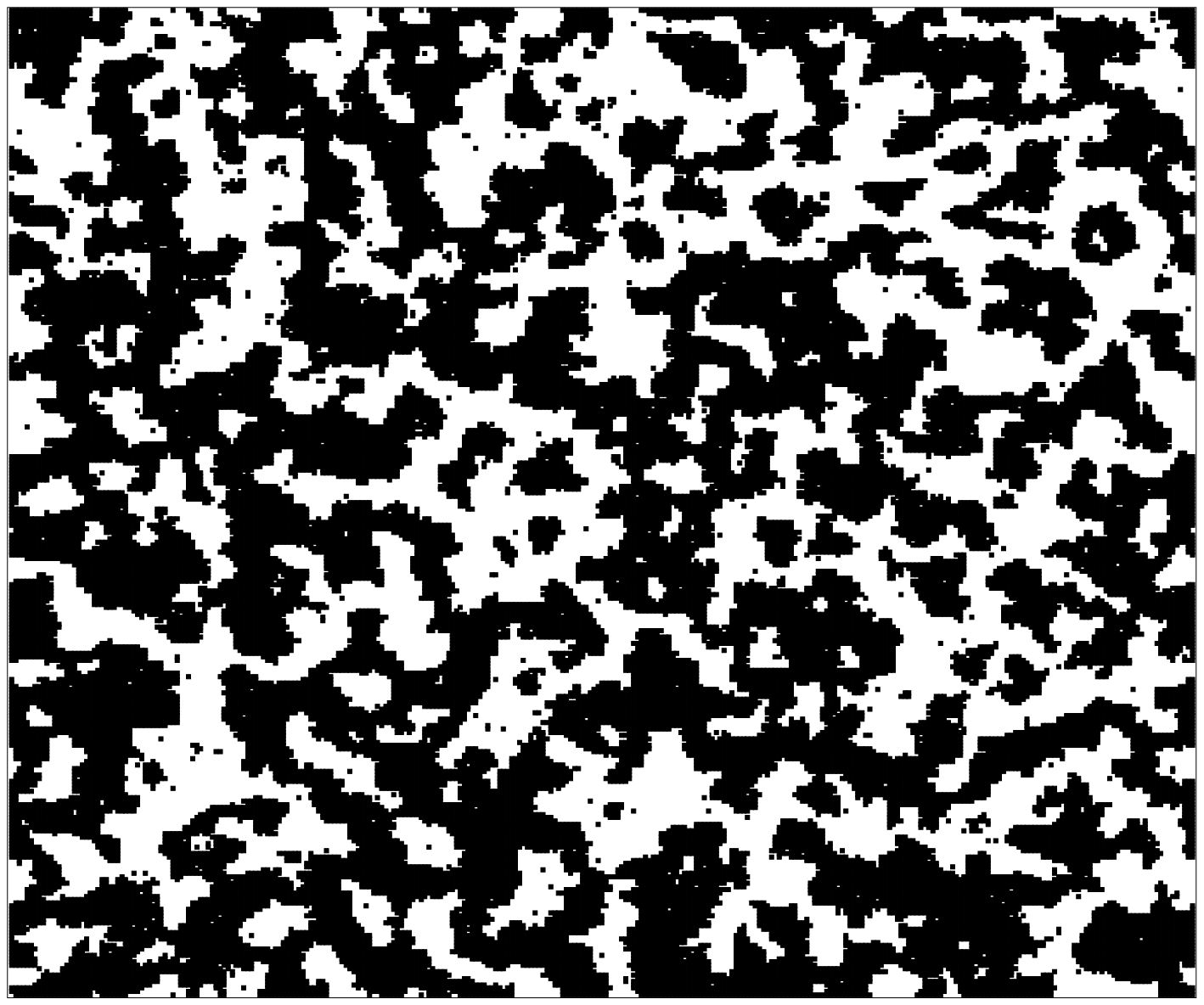} ~~~~~~~~~
\epsfxsize=2.1in\epsfbox{
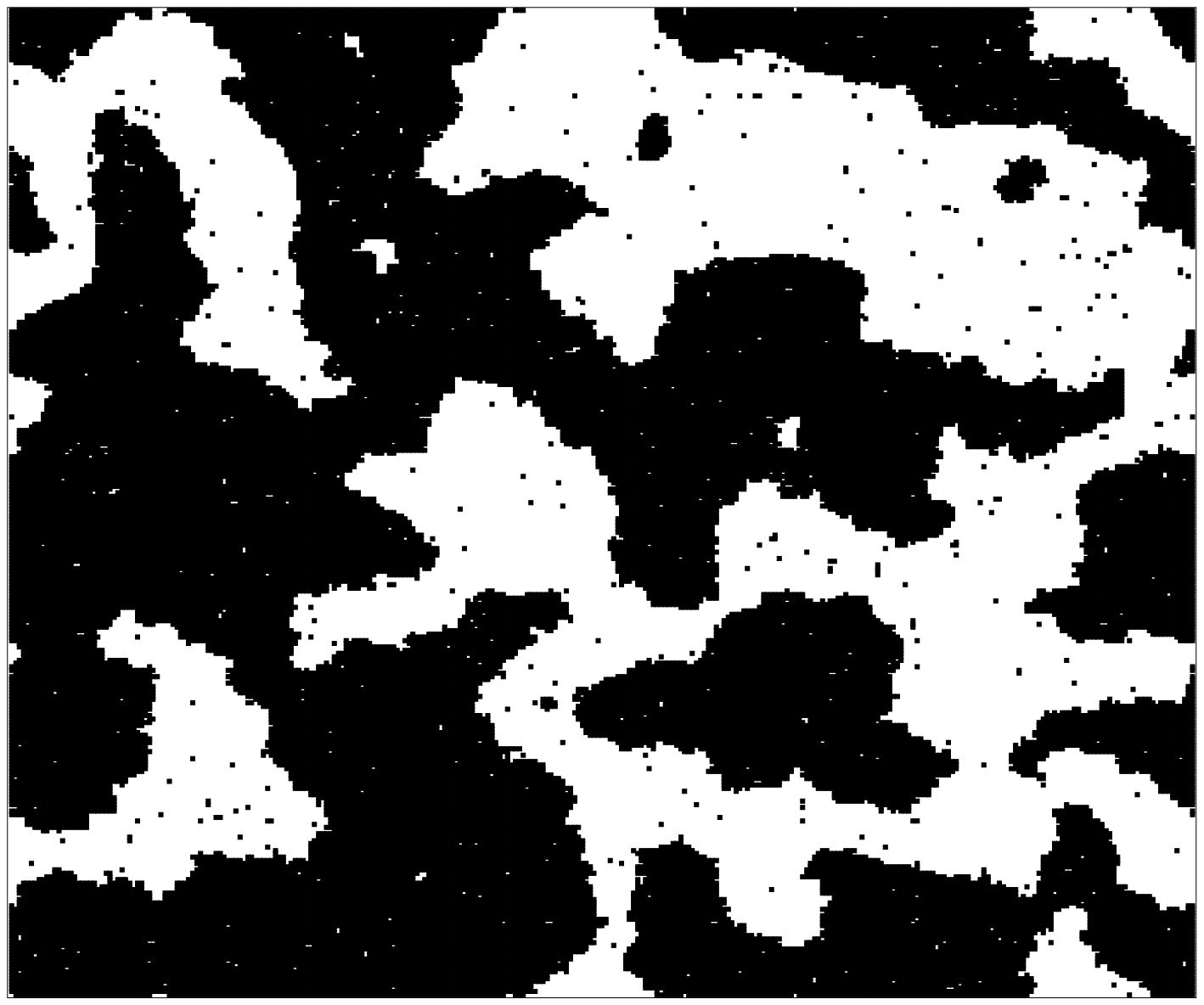}
}
\caption[Coarsening]{Snapshot of the coarsening of ordered domains in the 
$2D$ Glauber-Ising model, after a quench to $T=1.5<T_c$ from a totally 
disordered state and at times $t=25$ (left) and $t=275$ (right) after the 
quench. The black and white colours indicate the value of the Ising spins. 
\label{Bild1}}
\end{figure}

A typical example of the microscopic evolution of such a system is shown in
figure~\ref{Bild1}. Quite rapidly, ordered domains are formed which slowly
move and grow. Empirically, it is found that the typical size of the domains
scales with the time $t$ as $L(t)\sim t^{1/z}$, where $z$ is the dynamical
exponent. It is known that for dynamical rules chosen such that there is
no macroscopic conservation law, $z=2$ for quenches to $T<T_{\rm c}$. The slow
temporal evolution of macroscopic observables we are interested in results from 
the slow motion of the domain walls. It is common to consider
the coarse-grained order parameter (e.g. mean magnetization for magnetic 
systems) $\phi(t,\vec{r})$ and one tries to capture its time-evolution through 
a stochastic Langevin equation
\begin{equation} \label{Henkel:gl1-1}
\frac{\partial \phi}{\partial t} = - \frac{\delta {\cal H}[\phi]}{\delta \phi}
+\eta
\end{equation}
where the gaussian white noise $\eta=\eta(t,\vec{r})$ is characterized by its 
first two moments $\langle \eta(t,\vec{r})\rangle=0$ and 
$\langle \eta(t,\vec{r})\eta(t',\vec{r}')\rangle = 
2T\delta(t-t')\delta(\vec{r}-\vec{r}')$ and ${\cal H}$ is the Ginzburg-Landau 
functional. For systems undergoing a conventional second-order 
phase-transition, one expects that qualitatively 
\begin{equation} \label{Henkel:gl1-2}
{\cal H}[\phi] = \phi \Delta \phi + {\cal V}[\phi] \;\; ; \;\;
{\cal V}[\phi] \sim \left\{ \begin{array}{ll}
\phi^2 & \mbox{\rm ~~;~~ if~ $T>T_{\rm c}$} \\
\left( \phi^2-\phi_0^2 \right)^2 & \mbox{\rm ~~;~~ if~ $T<T_{\rm c}$}
\end{array} \right.
\end{equation}
where $\phi_0=\phi_0(T)$ are the two equilibrium values of $\phi$ and $\Delta$
is the spatial Laplacian. 
Physically, systems with $T>T_{\rm c}$ and $T<T_{\rm c}$ are very different,
since in the first case, there is a single ground-state (where ${\cal H}[\phi]
= \min !$) while there are two distinct ground states in the second case. 
Therefore, for $T>T_{\rm c}$ the system will rapidly relax towards it single
ground-state and no ageing occurs. On the other hand, if $T<T_{\rm c}$ it will
depend on the microscopic environment of each spin variable to which of the
two possible local ground-states $\pm \phi_0$ the system will evolve locally. 
The competition between these distinct states then leads to ageing phenomena. 

The noisy Langevin equation can be turned into an equivalent field theory 
through the Martin-Siggia-Rose formalism, see e.g. \cite{Henkel:Cugl02}. 
Schematically, it may be represented through the effective action
\begin{equation} \label{Henkel:gl1-3}
S[\phi,\wit{\phi}] = \int \!{\rm d} t\,{\rm d}\vec{r}\: 
\left[ \wit{\phi}\left(\frac{\partial \phi}{\partial t}+
\frac{\delta {\cal H}[\phi]}{\delta \phi}\right)
-T \wit{\phi}^2 \right]
\end{equation}
where $\wit{\phi}$ is the so-called response field conjugate to $\phi$. From
this, the classical equations of motion take the form
\begin{equation} \label{Henkel:gl1-4}
\frac{\partial\phi}{\partial t} = - \frac{\delta {\cal H}}{\delta \phi} 
+2 T \wit{\phi} \;\; ; \;\;
\frac{\partial\wit{\phi}}{\partial t} =
\frac{\delta^2 {\cal H}}{\delta\phi^2}\,\wit{\phi}
\end{equation}
but initial conditions must still be specified. For interacting fields,
fluctuation effects are not taken into account by the equations 
(\ref{Henkel:gl1-4}), although they are present in the action $S$ and the 
associated path integral. It turns out, see \cite{Henkel:Godr02} for a 
review, that ageing is more 
fully revealed through the study of two-time correlators $C(t,s)$ and response 
functions $R(t,s)$ defined by
\begin{equation}
C(t,s) = \langle \phi(t,\vec{r})\phi(s,\vec{r})\rangle 
\;\; ; \;\;
R(t,s) = \left.\frac{\delta \langle\phi(t,\vec{r})\rangle}{\delta h(s,\vec{r})}
\right|_{h=0} = \langle \phi(t,\vec{r})\wit{\phi}(s,\vec{r})\rangle
\end{equation}
where $h(s,\vec{r})$ is the local magnetic field at time $s$ and the position
$\vec{r}$. The last equation comes from Martin-Siggia-Rose theory. 
Furthermore, $\langle \wit{\phi}\wit{\phi}\rangle=0$.  

\begin{definition}
A statistical system described by a Langevin equation (\ref{Henkel:gl1-1}) or 
an effective action (\ref{Henkel:gl1-3}) is said to undergo {\em ageing}, if
$C=C(t,s)$ or $R=R(t,s)$ depend on both the {\em observation time} $t$ and 
the {\em waiting time} $s$ and not merely on the difference $\tau=t-s$.
\end{definition}

If the times $t$, $s$ and $t-s$ become large simultaneously (as compared to 
some microscopic time scale $t_{\rm micro}$), one usually finds, 
for $T<T_{\rm c}$, the following scaling behaviour, 
see e.g. \cite{Henkel:Godr02}
\begin{eqnarray}
C(t,s) \simeq M_{\rm eq}^2 f_C(t/s) &;& f_C(x) \sim x^{-\lambda_C/z} 
\mbox{\rm ~~ ,~ $x\to\infty$} 
\nonumber \\
R(t,s) \simeq s^{-1-a} f_R(t/s) &;& f_R(x) \sim x^{-\lambda_R/z} 
\mbox{\rm ~~ ,~ $x\to\infty$}
\label{Henkel:gl1-6}
\end{eqnarray}
where $M_{\rm eq}$ is the equilibrium magnetization and $\lambda_{C,R}$ are the
autocorrelation \cite{Henkel:Fish88} and autoresponse \cite{Henkel:Pico02} 
exponents, respectively. If long-range initial correlations of the form 
$C_{\rm ini}(\vec{r})\sim |\vec{r}|^{-d-\alpha}$ are used, where $d$ is the 
number of space dimensions and $\alpha$ a free parameter, one has for
{\em ferro}magnets the rigorous bound $\lambda_C\geq (d+\alpha)/2$ 
\cite{Henkel:Yeun96} (it need not hold for disordered systems 
\cite{Henkel:Sche03}). Furthermore,
if $\alpha<0$, the relationship $\lambda_C=\lambda_R+\alpha$ has been 
conjectured \cite{Henkel:Pico02}, while $\lambda_C=\lambda_R$ for a fully 
disordered initial state is generally accepted. Finally, the value of the 
exponent $a$ has recently been shown \cite{Henkel:Henk02a} by scaling arguments 
to depend on the equilibrium spin-spin correlator $C_{\rm eq}$ as follows. 
Systems of {\em class S} have short-ranged correlators 
$C_{\rm eq}(\vec{r})\sim e^{-|\vec{r}|/\xi}$ and systems of {\em class L} 
have long-ranged correlators $C_{\rm eq}(\vec{r})\sim |\vec{r}|^{-(d-2+\eta)}$.
Then \cite{Henkel:Henk02a}
\begin{equation}
a = \left\{ \begin{array}{ll}
1/z & \mbox{\rm ~~;~ class S} \\
(d-2+\eta)/z & \mbox{\rm ~~;~ class L}
\end{array} \right.
\end{equation}
This concludes our review of those main properties of ageing systems which we 
shall need below. 

\section{Beyond scale invariance} 
 
Our central question about ageing is the following: 
Is there a general, model-independent way to predict the form of the scaling
functions $f_C(x)$ and $f_R(x)$ as defined in (\ref{Henkel:gl1-6})~? 

The possibility of an affirmative answer might be suggested by the 
following known facts: 
(i) Ageing phenomena show a dynamical scaling, i.e. they are scale-invariant. 
(ii) In equilibrium critical phenomena, the invariance of the theory under
dilatations $\vec{r}\mapsto b \vec{r}$ may be extended to local scale
or conformal transformations $\vec{r}\mapsto b(\vec{r}) \vec{r}$ (such that
angles are kept fixed). It is well-known that conformal invariance is a
very powerful principle in two-dimensional equilibrium critical phenomena,
see e.g. \cite{Henkel:Henk99}. 
More precisely, we inquire whether the global dynamical
scale transformations $t\mapsto b^z t$ and $\vec{r}\mapsto b \vec{r}$ can be
extended, analogously to conformal invariance where $z=1$, to space-time 
dependent dilatation factors $b=b(t,\vec{r})$ in a meaningful way. 

\begin{example}
Let $z=2$ and consider $d$ space dimensions. The {\em Schr\"odinger group} 
{\sl Sch}($d$) is defined by \cite{Henkel:Nied72}
\begin{equation}
t\mapsto \frac{\alpha t+\beta}{\gamma t +\delta} \;\; , \;\;
\vec{r}\mapsto \frac{{\cal R}\vec{r}+\vec{v}t+\vec{a}}{\gamma t +\delta} 
\;\; ; \;\;
\alpha\delta-\beta\gamma=1
\end{equation}
where ${\cal R}\in SO(d)$, $\vec{a},\vec{v}\in\mathbb{R}^d$ and 
$\alpha,\beta,\gamma,\delta\in\mathbb{R}$. It is well-known that {\sl Sch}($d$) 
is the maximal kinematic group of 
the free Schr\"odinger equation ${\cal S}\psi=0$ with 
${\cal S}=2m{\rm i}\partial_t - \partial_{\vec{r}}^2$ \cite{Henkel:Nied72} 
(that is, it maps any solution of ${\cal S}\psi=0$ to another solution). 
Additional examples of {\sl Sch}($d$) as a kinematic group include certain 
non-linear Schr\"odinger equations \cite{Henkel:Niki03}, 
systems of Schr\"odinger equations \cite{Henkel:Cher99} and
the Euler equations of fluid dynamics \cite{Henkel:ORai01}. We denote the 
Lie algebra of {\sl Sch}($d$) by $\mathfrak{sch}_d$. Specifically,
$\mathfrak{sch}_1=\overline{\{X_{\pm 1,0}, Y_{\pm 1/2}, M_0\}}$ with the 
non-vanishing commutation relations
\begin{equation}
\left[ X_{n}, X_{n'}\right] = (n-n') X_{n+n'} \;\; , \;\;
\left[ X_{n}, Y_{m}\right] = \left(\frac{n}{2}-m\right) Y_{n+m} \;\; , \;\;
\left[ Y_{1/2}, Y_{-1/2}\right] = M_0
\end{equation}
where $n,n'\in\{\pm 1, 0\}$ and $m\in\{\pm 1/2\}$. 
\end{example}

\begin{example}
For a more general dynamical exponent $z\ne 2$, we construct infinitesimal
generators of local scale transformations from the following requirements
\cite{Henkel:Henk02} (for simplicity, set $d=1$): (a) Transformations in time 
are $t\mapsto(\alpha t+\beta)/(\gamma t+\delta)$ with 
$\alpha\delta-\beta\gamma=1$.
(b) The generator for dilatations is $X_0=-t\partial_t-z^{-1}r\partial_r-x/z$,
where $x$ is the scaling dimension of the fields $\phi,\wit{\phi}$ on which
the generators act. (c) Space-translation invariance is required, with
generator $-\partial_r$. Starting from these conditions, we can show by
explicit construction that there exist generators $X_n$, $n\in\{\pm 1, 0\}$ and
$Y_m$, $m=-1/z, 1-1/z,\ldots$ such that
\begin{equation} \label{Henkel:gl2-10}
\left[ X_{n}, X_{n'}\right] = (n-n') X_{n+n'} \;\; , \;\;
\left[ X_{n}, Y_{m}\right] = \left(\frac{n}{z}-m\right) Y_{n+m} 
\end{equation}
For generic values of $z$, it is sufficient to specify the `special' generator
\begin{equation} \label{Henkel:gl2-11}
X_1 = -t^2\partial_t -Ntr\partial_r -Nxt - \wit{\alpha}r^2\partial_t^{N-1}
-\wit{\beta} r^2\partial_r^{2(N-1)/N} -\wit{\gamma}\partial_r^{2(N-1)/N} r^2
\end{equation}
explicitly, where we wrote $z=2/N$ and $\wit{\alpha},\wit{\beta},\wit{\gamma}$ 
are free constants. Further non-generic solutions exist for $N=1$ and $N=2$ 
\cite{Henkel:Henk02}. In particular, the generator (\ref{Henkel:gl2-11}) 
reproduces for $z=2$ those of {\sl Sch}(1). The condition $[X_1, Y_{N/2}]=0$ 
is only satisfied if either 
(I) $\wit{\beta}=\wit{\gamma}=0$ which we call {\em type I} or else 
(II) $\wit{\alpha}=0$ which we call {\em type II}. 
\end{example}
\begin{definition} 
If a statistical system is invariant under the infinitesimal generators of
either type I or type II it is said to be {\em locally scale-invariant} of
type I or type II, respectively. 
\end{definition}
\noindent 
Only the generators of type II are suitable for applications to ageing
phenomena. 
\begin{theorem} The generators $X_n, Y_m$ of type II form a kinematic symmetry 
of the differential equation ${\cal S}\psi=0$ where
\begin{equation}
{\cal S} = -(\wit{\beta}+\wit{\gamma})\partial_t + \frac{1}{z^2}\partial_r^z
\end{equation}
\end{theorem}
\noindent 
For $z=2$, we recover the $d=1$ free Schr\"odinger equation and its maximal 
kinematic Lie algebra $\mathfrak{sch}_1$. See \cite{Henkel:Henk02} for the
precise definition of the commuting fractional derivatives $\partial_r^z$. 

In order to be able to apply this kinematic symmetry to ageing phenomena, 
we must consider the subset of $\{X_n,Y_m\}$ where time translations
(generated by $X_{-1}=-\partial_t$) are left out. It can be checked that the 
initial line $t=0$ is kept invariant. Ageing systems are not time-translation 
invariant and local scale invariance for them is meant to exclude the 
generator $X_{-1}$. 
\begin{theorem} \label{Henkel:th2}
Consider an ageing statistical system which is locally scale-invariant of
type II. Then the autoresponse function is
\begin{equation} \label{Henkel:gl2-13}
R(t,s) = r_0\left(\frac{t}{s}\right)^{1+a-\lambda_R/z}\left(t-s\right)^{-1-a}
\;\; ; \;\; t>s
\end{equation}
where $r_0$ is a normalization constant. Furthermore, consider the following
scaling form of the spatio-temporal response
\begin{equation} \label{Henkel:gl2-14}
R(t,s;\vec{r}) = \left.
\frac{\delta\langle\phi(t,\vec{r})\rangle}{\delta h(s,\vec{0})}
\right|_{h=0} = R(t,s) \Phi\left(\frac{r}{(t-s)^{1/z}}\right)
\end{equation}
Then $\Phi(u)$ is a solution of the equation
\begin{equation} \label{Henkel:gl2-15}
\left( \partial_u +z\left(\wit{\beta}+\wit{\gamma}\right)u\partial_u^{2-z}
+2z(2-z)\wit{\gamma}\partial_u^{1-z}\right) \Phi(u) = 0
\end{equation}
In the special case $z=2$, we have
\begin{equation} \label{Henkel:gl2-16}
R(t,s;\vec{r}) = R(t,s)\exp\left(-\frac{\cal M}{2}\frac{\vec{r}^2}{t-s}\right)
\end{equation}
where ${\cal M}=\wit{\beta}+\wit{\gamma}$ is a constant. 
\end{theorem}
\begin{proof}
\cite{Henkel:Henk01b,Henkel:Henk02} 
Consider first the autoresponse $R=R(t,s)=\langle\phi\wit{\phi}\rangle$, where
$\phi,\wit{\phi}$ have the scaling dimensions $x_{\phi},x_{\wit{\phi}}$,
respectively. Local scale invariance means that
$X_n R = Y_m R=0$, with $n\geq 0$. Because of spatial translation invariance 
and the commutators (\ref{Henkel:gl2-10}), it is sufficient to check that
$X_0 R = X_1 R=0$. The explicit form (\ref{Henkel:gl2-11}) produces two 
linear differential equations for $R(t,s)$ which are readily solved. 
Comparison with the scaling forms (\ref{Henkel:gl1-6}) then 
establishes (\ref{Henkel:gl2-13}). Eq.~(\ref{Henkel:gl2-15}) is proven
similarly, see \cite{Henkel:Henk02}. Eq.~(\ref{Henkel:gl2-16}) had been found
earlier \cite{Henkel:Henk94}. 
\end{proof}
In these explicit forms of $R(t,s;\vec{r})$ it is also assumed that the 
system is rotation-invariant. If that is not the case, $\cal M$ is no longer
uniform, but becomes direction-dependent \cite{Henkel:Henk02b}. 

Eq.~(\ref{Henkel:gl2-13}) has been tested and confirmed in several spin systems
undergoing ageing, notably the kinetic Ising model with Glauber dynamics
in $d=2,3$ through intensive simulations \cite{Henkel:Henk01b,Henkel:Henk02a} 
and the exactly solvable kinetic spherical model in $d>2$ dimensions 
\cite{Henkel:Cann01,Henkel:Godr02,Henkel:Pico02}, for $T\leq T_{\rm c}$, the 
random walk \cite{Henkel:Cugl94} and finally (up to logarithmic correction 
factors) the $2D$ XY model with $T<T_{\rm c}$ \cite{Henkel:Abri03} and the
$2D$ critical voter model \cite{Henkel:Sast03}. 
Furthermore, eq.~(\ref{Henkel:gl2-16}) has been numerically confirmed in the 
Glauber-Ising model, again for $d=2,3$ and $T<T_{\rm c}$ \cite{Henkel:Henk02b}. 
However, small corrections with respect to (\ref{Henkel:gl2-13}) were found 
at $T=T_{\rm c}$ in a two-loop $\varepsilon$-expansion \cite{Henkel:Cala02} 
and in a recent self-consistent study \cite{Henkel:Maze03} of the 
time-dependent Ginzburg-Landau equation at $T=0$ and which improves on the 
approximate Ohta-Jasnow-Kawasaki theory. 
We refer to the literature for details. 

\section{On the Schr\"odinger group}

Having reviewed the important phenomenological result of local scale 
invariance as given in Theorem~\ref{Henkel:th2}, we now discuss in more detail
how a dynamical symmetry such as local scale invariance might come about. 
We shall do this here for the case $z=2$ and therefore discuss the Schr\"odinger
group and Schr\"odinger-invariant systems \cite{Henkel:Henk03}. 
For simplicity, we often set $d=1$. 

Under the action of an element $g\in\mbox{\sl Sch}(1)$ of the Schr\"odinger 
group, a solution $\phi(t,r)$ of the free equation 
$(2{\cal M}\partial_t - \partial_r^2)\phi=0$ 
where ${\cal M}={\rm i} m$ is fixed, 
transforms projectively, viz. 
$\phi(t,r)\mapsto  (T_g\phi)(t,r)=f_g(g^{-1}(t,r))\phi(g^{-1}(t,r))$ with a
known \cite{Henkel:Nied72} companion function $f_g$. 

\begin{figure}[t]
\centerline{\epsfxsize=1.5in\ \epsfbox{
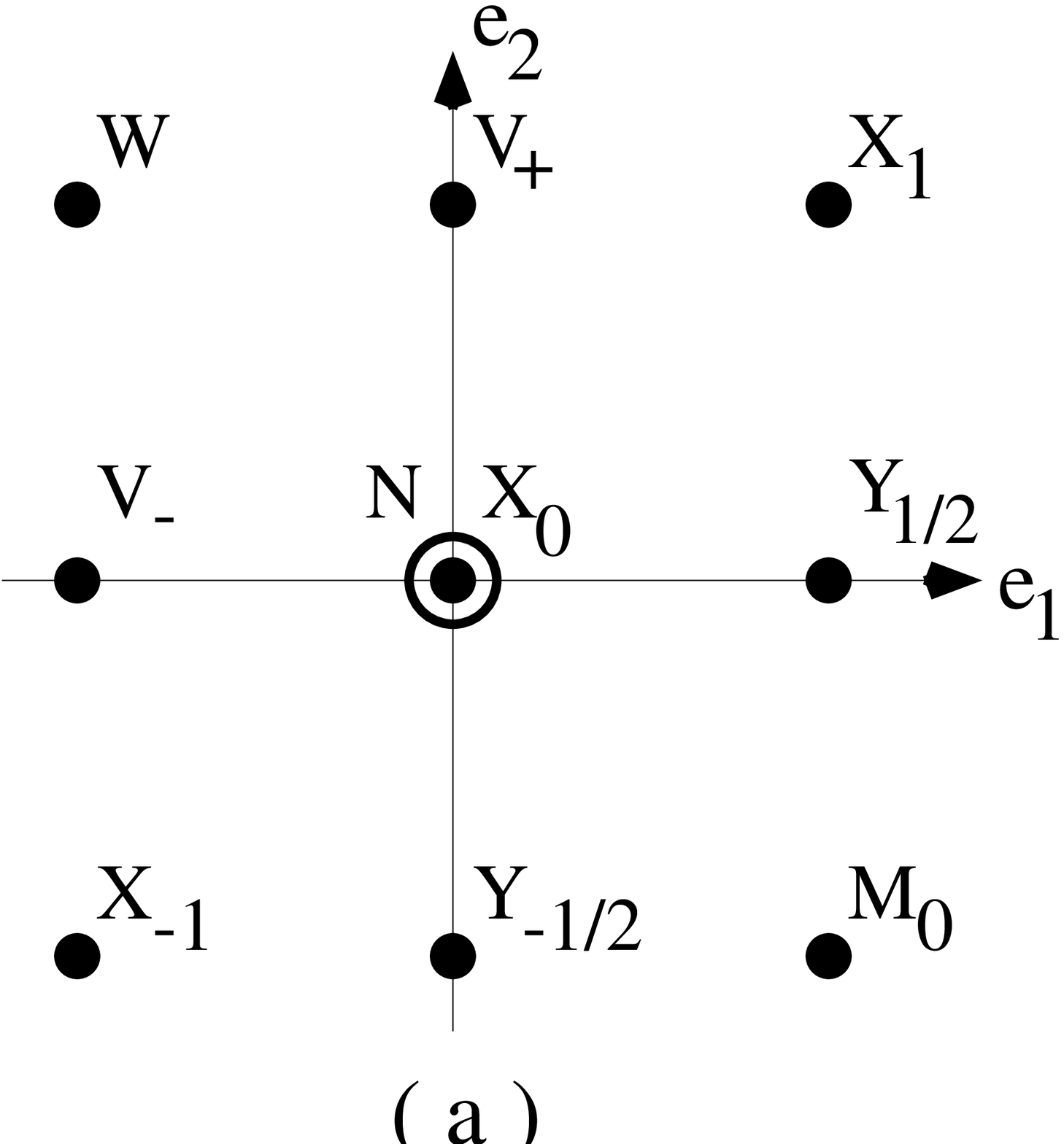} ~
\epsfxsize=1.5in\epsfbox{
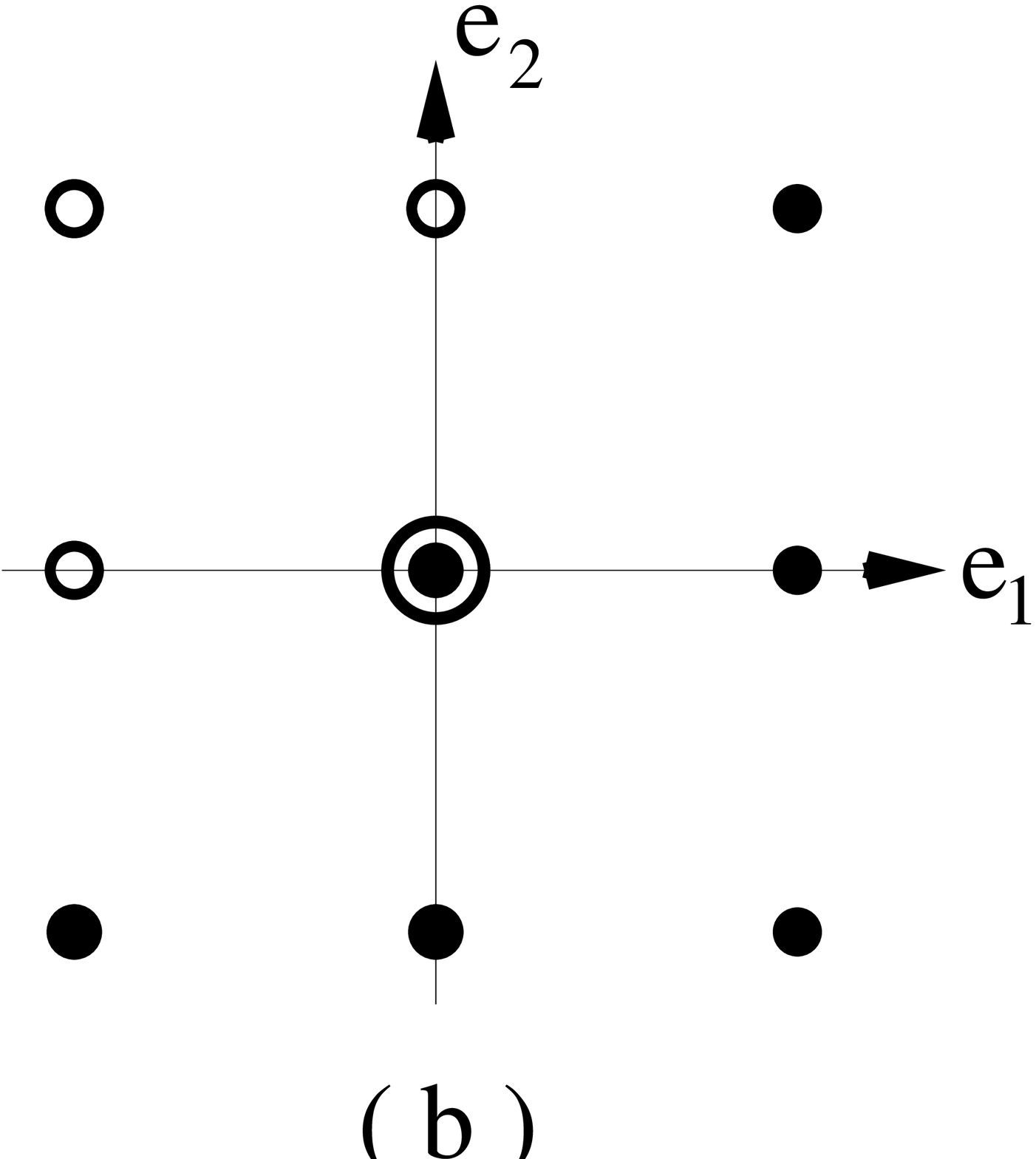} ~
\epsfxsize=1.5in\epsfbox{
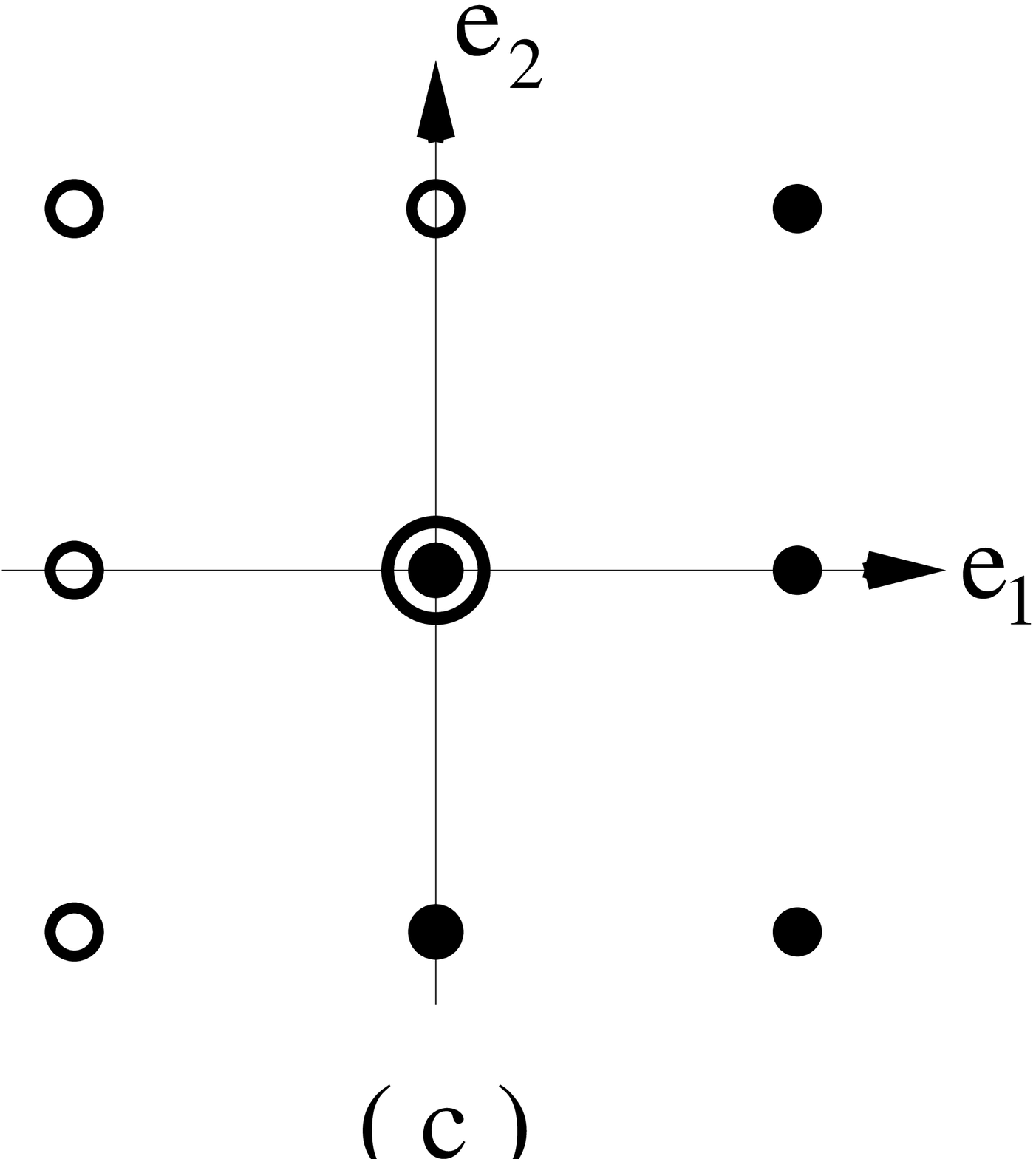} ~
\epsfxsize=1.5in\epsfbox{
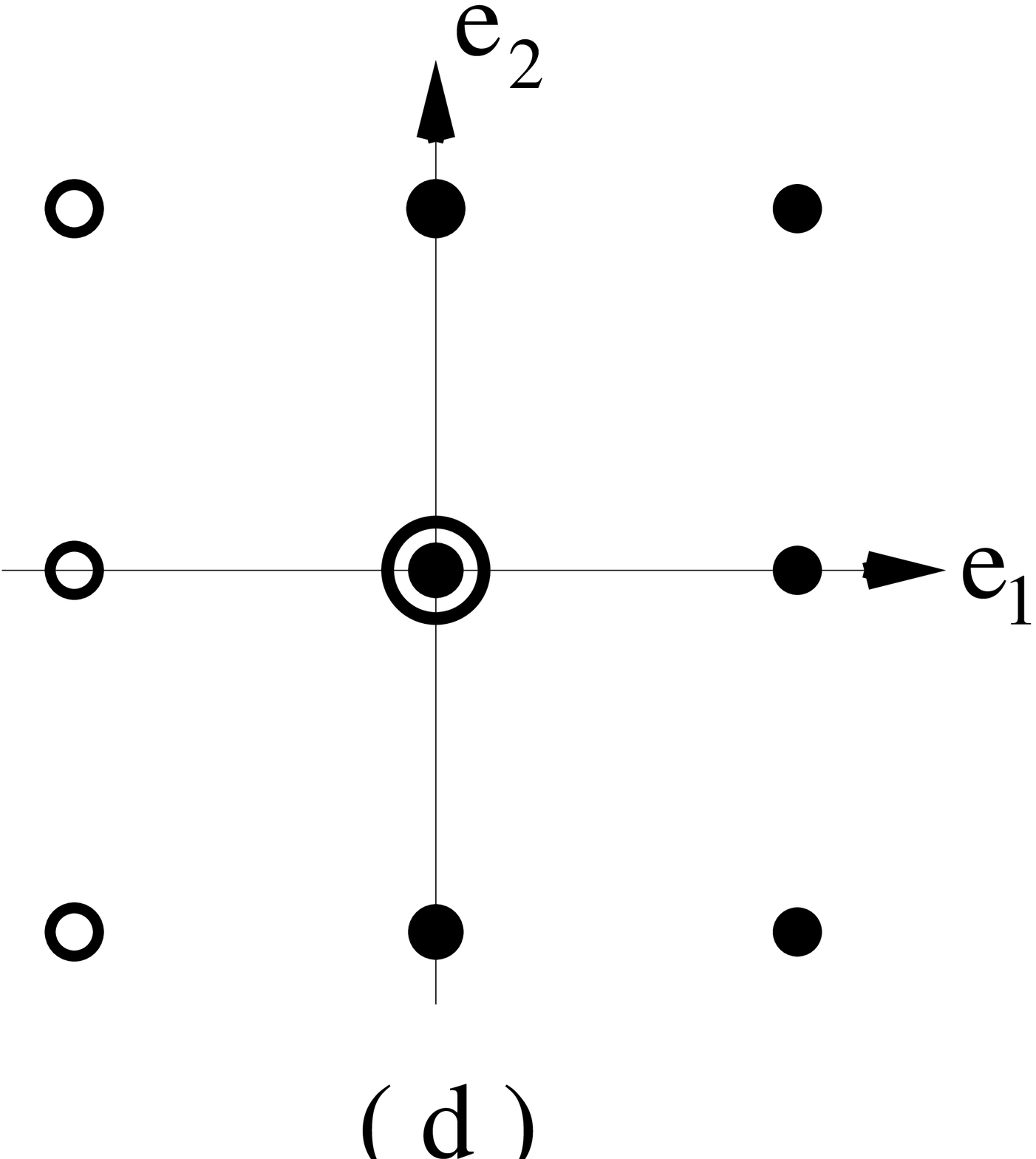}
}
\caption[Root space]{(a) Roots of the complex Lie algebra $B_2$ and the 
identification of the generators of the complexified conformal Lie algebra
$(\mathfrak{conf}_3)_{\mathbb{C}}\supset(\mathfrak{sch}_1)_{\mathbb{C}}$.
The double circle in the center denotes the Cartan subalgebra.   
The generators belonging to the three non-isomorphic parabolic subalgebras 
\protect{\cite{Henkel:Henk03}} 
are indicated by the full points, namely
(b) $\wit{\mathfrak{sch}}_1$, (c) $\wit{\mathfrak{age}}_1$ and
(d) $\wit{\mathfrak{alt}}_1$. 
\label{Bild2}}
\end{figure}

Following \cite{Henkel:Giul96}, we treat $\cal M$ as an additional variable
and ask about the maximal kinematic group in this case 
\cite{Henkel:Henk03}. First introduce a new coordinate $\zeta$ and a 
new wave function $\psi$ through
\begin{equation} \label{Henkel:gl3-17}
\phi(t,r) = \frac{1}{\sqrt{2\pi}} \int_{\mathbb{R}} \!{\rm d}\zeta\, 
e^{-{\rm i}{\cal M}\zeta} \psi(\zeta,t,r)
\end{equation}
We denote time $t$ as the zeroth coordinate and 
$\zeta$ as coordinate number $-1$. 
When working out the action of the generators of $\mathfrak{sch}_1$ on the
function $\psi$, it is easily seen that the projective phase factors can be
absorbed into certain translations of the variable $\zeta$ \cite{Henkel:Henk03}.
Furthermore, the free Schr\"odinger equation becomes
\begin{equation} \label{Henkel:gl3-18}
\left( 2{\rm i} \frac{\partial^2}{\partial\zeta \partial t} + 
\frac{\partial^2}{\partial r^2} \right) \psi(\zeta,t,r) = 0
\end{equation}
In order to find the maximal kinematic symmetry of this equation, we recall that
the three-dimensional Klein-Gordon equation 
$\sum_{\mu=-1}^{1}\partial_{\mu}\partial^{\mu} \Psi(\vec{\xi})=0$ has the
$3D$ conformal algebra $\mathfrak{conf}_3 \cong so(4,1) \cong B_2$ as maximal
kinematic symmetry. By making the following change of variables
\begin{equation}
\zeta = \frac{1}{2} \left( \xi_0 + {\rm i} \xi_{-1} \right) \;\; , \;\;
t = \frac{1}{2} \left( -\xi_0 + {\rm i} \xi_{-1} \right) \;\; , \;\;
r = \sqrt{\frac{{\rm i}}{2}}\, \xi_1
\end{equation}
and setting $\psi(\zeta,t,r)=\Psi(\vec{\xi})$, the $3D$ Klein-Gordon equation
reduces to (\ref{Henkel:gl3-18}). Therefore \cite{Henkel:Henk03}
\begin{theorem} \label{Henkel:th3}
For variable masses $\cal M$, the maximal kinematic symmetry algebra of the 
free Schr\"odinger equation in $d$ dimensions is isomorphic to the conformal 
algebra $\mathfrak{conf}_{d+2}$ and one has the inclusion of the complexified
Lie algebras
$(\mathfrak{sch}_d)_{\mathbb{C}} \subset (\mathfrak{conf}_{d+2})_{\mathbb{C}}$.
\end{theorem}
For $d=1$, the Cartan subalgebra is spanned by the generators $X_0$ and 
$N:=-t\partial_t + \zeta\partial_{\zeta}$. With respect to the six generators 
of the Schr\"odinger algebra $\mathfrak{sch}_1$, there are the four additional
ones $N,V_+, V_-, W$. They are identified in figure~\ref{Bild2}a with the 
roots in the root space of $B_2$. 

Consider the non-isomorphic parabolic subalgebras of $B_2$. 
In figure~\ref{Bild2}, these correspond to convex subsets of roots. There are
two maximal parabolic subalgebras, namely 
(i) $\wit{\mathfrak{sch}}_1 := \mathfrak{sch}_1 \oplus \mathbb{C}N$ and 
(ii) $\wit{\mathfrak{alt}}_1 := \mathfrak{alt}_1 \oplus \mathbb{C}N$ where
$\mathfrak{alt}_1 := \overline{\{X_0,X_1, Y_{\pm 1/2}, M_0, V_+\}}$ and
$V_+=-2\zeta r\partial_{\zeta}-
2tr\partial_t-(r^2+2{\rm i} \zeta t)\partial_r-2xr$, 
see also figure~\ref{Bild2}bd. The minimal parabolic subalgebra is
$\wit{\mathfrak{age}}_1 := \mathfrak{age}_1 \oplus \mathbb{C}N$, where
$\mathfrak{age}_1 := \overline{\{X_0,X_1, Y_{\pm 1/2}, M_0\}}$, 
see figure~\ref{Bild2}c. We see that both $\wit{\mathfrak{alt}}_1$ and 
$\wit{\mathfrak{age}}_1$ do not
contain the generator $X_{-1}$ of time translations and they may therefore
be considered candidates for a dynamical symmetry of ageing systems. 

In writing down above the Klein-Gordon equation, we had used units such that
the `speed of light' $c=1$. It had been claimed in the literature that in 
a non-relativistic limit $c\to\infty$, there were a group contraction
$\mathfrak{conf}_{d+1}\curvearrowright\mathfrak{sch}_d$ provided that an 
ill-defined {\it ``\ldots transfer of the transformation of mass to 
the coordinates \ldots''} is carried out \cite{Henkel:Baru73}. 
However, going over first to the variable $\zeta$ before letting $c\to\infty$, 
we do not find a group contraction but rather the map 
$\mathfrak{conf}_3\to\wit{\mathfrak{alt}}_1\ne\mathfrak{sch}_1$.

We now discuss some physical consequences of Theorem~\ref{Henkel:th3}. 
Consider the effective action $S=S[\psi(\zeta,t,r)]$. 
For a free field $\psi$ one recovers from the action
\begin{equation} \label{Henkel:gl3-21}
S = \int \!{\rm d}\zeta\,{\rm d} t\,{\rm d} r\: \left[ 2{\rm i} 
\frac{\partial \psi}{\partial \zeta}\frac{\partial \psi}{\partial t}
+\left(\frac{\partial \psi}{\partial r}\right)^2\right] + S_{\rm ini}
\end{equation}
where $S_{\rm ini}$ describes the initial conditions, the Schr\"odinger 
equation (\ref{Henkel:gl3-18}) as equation of motion. For an infinitesimal
coordinate transformation parameterized by $\varepsilon_{\nu}$, $\nu=-1,0,1$, 
a theory is said to be {\em local} if its action $S$ transforms as 
(the second integral is restricted to the line $t=0$)
\begin{equation} \label{Henkel:gl3-22}
\delta S = \int \!{\rm d}\zeta\,{\rm d} t\,{\rm d} r\: 
{T_{\mu}}^{\nu}\partial^{\mu}\varepsilon_{\nu}
+\int_{(t=0)} \!{\rm d}\zeta\,{\rm d} r\: U^{\nu}\varepsilon_{\nu}
\end{equation}
where ${T_{\mu}}^{\nu}$ is the energy-momentum tensor . 
\begin{theorem} \label{Henkel:th4}
If the action $S$ of a local theory is invariant under translations in
$\zeta$ and $r$, scale-invariant with $z=2$ and Galilei-invariant, 
then $\delta_{X_1} S=0$ and $S$ in invariant under the action of 
$\mathfrak{age}_1$. 
\end{theorem}
\begin{proof} \cite{Henkel:Henk03} 
The line $t=0$ is invariant under the action of 
$\mathfrak{age}_1$. Furthermore, locality (\ref{Henkel:gl3-22}) yields several 
Ward identities. First, translation-invariance in $\zeta$ and $r$ implies 
$U^{-1}=U^{1}=0$. Dilatation invariance gives ${T_0}^0+\frac{1}{2}{T_1}^1=0$ 
and Galilei-invariance implies ${T_0}^1-{\rm i} {T_1}^{-1}=0$. Now, for an
infinitesimal special Schr\"odinger transformation, 
\begin{displaymath}
\delta_{X_1}S = -\varepsilon \int \!{\rm d}\zeta\,{\rm d} t\,{\rm d} r\: 
\left[ \left( 2{T_0}^0+{T_1}^1\right) t
+\left( {T_0}^1-{\rm i} {T_1}^{-1}\right) r\right]
+\frac{{\rm i} \varepsilon}{2} \int_{(t=0)} \!{\rm d}\zeta\,{\rm d} r\: 
r^2 U^{-1} = 0
\end{displaymath}
because of the Ward identities derived above.
\end{proof}
For the free-field action (\ref{Henkel:gl3-21}), an energy-momentum tensor 
which satisfies these Ward 
identities can be explicitly constructed \cite{Henkel:Henk03}. We stress that
Galilei-invariance must be included as a hypothesis and one cannot 
expect to be able to derive it from weaker assumptions. On the other hand,
time-translation invariance is not required. 

Finally, we reconsider the derivation of the spatio-temporal response for
$z=2$ \cite{Henkel:Henk03}. A standard result of conformal invariance 
(see \cite{Henkel:Henk99}) gives together with Theorem~\ref{Henkel:th3}
\begin{equation} \label{Henkel:gl3-23}
\langle \psi_1(\zeta_1,t_1,r_1)\psi_2(\zeta_2,t_2,r_2)\rangle = 
\langle \Psi(\vec{\xi}_1)\Psi(\vec{\xi}_2)\rangle = 
\psi_0\, \delta_{x_1,x_2}\, t^{-x_1} 
\left(\zeta+\frac{{\rm i}}{2}\frac{r^2}{t}\right)^{-x_1}
\end{equation}
where $x_{1,2}$ are the scaling dimensions of the fields $\psi_{1,2}$, 
$\psi_0$ is a normalization constant and $\zeta=\zeta_1-\zeta_2$ and so on. 
Using the convention ${\cal M}_{1,2}\geq 0$, it turns out that 
eqs.~(\ref{Henkel:gl3-17},\ref{Henkel:gl3-23}) together imply that 
$\langle\phi_1\phi_2\rangle=0$.\footnote{This holds true for $T=0$. 
For the case $0<T<T_{\rm c}$, see \cite{Henkel:Pico03}.} However, if we define 
a {\em conjugate field} $\phi^*$ by formal complex conjugation in 
(\ref{Henkel:gl3-17}) with ${\cal M}\geq 0$, then for $x_1>0$ we find
\begin{eqnarray}
\langle \phi_1(t_1,r_1)\phi_2^*(t_2,r_2)\rangle &=& 
\frac{1}{2\pi} \int_{\mathbb{R}^2}\!{\rm d}\zeta_1 {\rm d}\zeta_2\: 
e^{-{\rm i}{\cal M}_1\zeta_1+{\rm i}{\cal M}_2\zeta_2} 
\langle \psi_1(\zeta_1,t_1,r_1)\psi_2(\zeta_2,t_2,r_2)\rangle
\nonumber \\
&=& \phi_0\, \delta_{x_1,x_2}\, \delta({\cal M}_1-{\cal M}_2)\, \Theta(t) 
t^{-x_1} \exp\left(-\frac{{\cal M}_1}{2}\frac{r^2}{t}\right)
\end{eqnarray}
and we stress that the $\Theta$-function expresses the causality condition
$t_1>t_2$ which is required for an interpretation of 
$\langle\phi_1\phi_2^*\rangle$ as response function. If we recall from
Martin-Siggia-Rose theory that 
$R(t,s;\vec{r})=\langle\phi(t,\vec{r})\wit{\phi}(s,\vec{0})\rangle=
\langle\phi(t,\vec{r})\phi^*(s,\vec{0})\rangle$, where effectively $\phi$ has
a `mass' ${\cal M}\geq 0$ and $\phi^*$ has a `mass' ${\cal M}^*\leq 0$, 
we have the identification
\begin{equation}
\phi^*(s,\vec{r}) = \wit{\phi}(s,\vec{r})
\end{equation}
This result has also been confirmed for the three-point response
functions \cite{Henkel:Henk03}. 

We now apply this to the parabolic subalgebras. For a system invariant under
$\mathfrak{age}_1$, we find
\begin{equation} \label{Henkel:gl3-26}
\langle \psi_1(\zeta_1,t_1,r_1)\psi_2(\zeta_2,t_2,r_2)\rangle = 
\psi_0  \left(\frac{t_1}{t_2}\right)^{(x_2-x_1)/2} t^{-(x_1+x_2)/2} 
\left(\zeta+\frac{{\rm i}}{2}\frac{r^2}{t}\right)^{-(x_1+x_2)/2}
\end{equation}
If we would consider the extension 
$\wit{\mathfrak{age}}_1\to\wit{\mathfrak{sch}}_1$, then invariance under
time translations generated by $X_{-1}=-\partial_t$ fixes $x_1=x_2$ and we
simply recover the result (\ref{Henkel:gl3-23}) coming form the invariance 
under $\mathfrak{conf}_3$. On the other hand, the extension 
$\wit{\mathfrak{age}}_1\to\wit{\mathfrak{alt}}_1$ requires that the 
condition $V_{+}\langle \psi_1\psi_2\rangle=0$ holds. It is easy to see from
(\ref{Henkel:gl3-26}) that this is the case. Therefore, for both 
$\wit{\mathfrak{age}}_1$ and $\wit{\mathfrak{alt}}_1$ we find, provided
$x_1+x_2>0$
\begin{equation}
\langle \phi_1\phi_2^*\rangle = \phi_0 \delta({\cal M}_1-{\cal M}_2) \Theta(t)
\left(\frac{t_1}{t_2}\right)^{(x_2-x_1)/2} t^{-(x_1+x_2)/2} 
\exp\left(-\frac{{\cal M}_1}{2}\frac{r^2}{t}\right)
\end{equation}
and we have indeed rederived (\ref{Henkel:gl2-16}) for ageing systems with
$z=2$ in a model-independent way, including the causality condition. 
There is not yet a criterion which would allow to decide if 
$\wit{\mathfrak{age}}_1$ or $\wit{\mathfrak{alt}}_1$, if any, is the dynamic
symmetry of ageing systems with $z=2$. 

In conclusion, the explicit form of the spatio-temporal two-time response 
function $R(t,s;\vec{r})$ as given in (\ref{Henkel:gl2-16}) is a consequence 
of the assumed Galilei-invariance of the ageing system. The high-precision 
numerical confirmation of this form in the kinetic Ising model with Glauber 
dynamics quenched to $T<T_c$ for $d=2$ and $d=3$ dimensions 
\cite{Henkel:Henk02b} is a strong indication in favour of Galilei-invariance 
in that model. However, even for $T=0$ the Langevin equation 
(\ref{Henkel:gl1-1}) or the system (\ref{Henkel:gl1-4}) are for general 
${\cal H}[\phi]$ not Galilei/Schr\"odinger-invariant, 
see \cite{Henkel:Cher99,Henkel:Niki03}. In view of small corrections to
(\ref{Henkel:gl2-13}) suggested by two-loop field-theory calculations
\cite{Henkel:Cala02,Henkel:Maze03} and of Theorem~\ref{Henkel:th4} on the 
other hand, it remains to be understood in what
precise sense ageing systems might be said to be Galilei-invariant.

\subsection*{Acknowledgements}

It is a pleasure to thank the organizers for the stimulating environment during
the conference and M. Pleimling and J. Unterberger for fruitful collaborations.

\LastPageEnding


\begin{thebibliography}{99}
\footnotesize

\bibitem{Henkel:Abri03}
S. Abriet and D. Karevski,
Off-equilibrium dynamics in the $2d$ XY system;
{\tt cond-mat/0309342}

\bibitem{Henkel:Baru73}
A.O. Barut, 
Conformal group $\to$ Schr\"odinger group $\to$ dynamical group: the maximal
kinematical group of the massive Schr\"odinger particle,
Helv. Phys. Acta {\bf 46}, 496--503 (1973). 

\bibitem{Henkel:Cala02} 
P. Calabrese and A. Gambassi,
Two-loop critical fluctuation-dissipation ratio for the relaxational dynamics
of the $O(N)$ Landau-Ginzburg hamiltonian,
Phys. Rev. {\bf E66}, 066101 (2002); {\tt cond-mat/0207452}.

\bibitem{Henkel:Cann01}
S.A. Cannas, D.A. Stariolo and F.A. Tamarit, 
Dynamics of ferromagnetic spherical spin models with power law interactions: 
exact solution,
Physica {\bf A294}, 362--374 (2001); {\tt cond-mat/0010319}. 

\bibitem{Henkel:Cate00} 
M.E. Cates and M.R. Evans (eds), 
{\it Soft and Fragile Matter}, 
Bristol, IOP Press (2000).

\bibitem{Henkel:Cher99} 
R. Cherniha and J.R. King, 
Lie symmetries of nonlinear multidimensional systems I \& II, 
J. Phys. {\bf A33}, 267--282 and 7839--7841 (2000); 
{\bf A36}, 405--425 (2003). 

\bibitem{Henkel:Cugl94}
L.F. Cugliandolo, J. Kurchan and G. Parisi, 
Off-equilibrium dynamics and ageing in unfrustrated systems, 
J. Physique {\bf I4}, 1641--1656 (1994); {\tt cond-mat/9406053}.

\bibitem{Henkel:Cugl02} 
L.F. Cugliandolo, 
Dynamics of glassy systems; 
{\tt cond-mat/0210312}

\bibitem{Henkel:Fish88} 
D.S. Fisher and D.A. Huse, 
Nonequilibrium dynamics of the spin-glass ordered phase, 
Phys. Rev. {\bf B38}, 373--385 (1988). 

\bibitem{Henkel:Giul96} 
D. Giulini, 
On Galilei invariance in quantum mechanics and the Bargmann superselection rule,
Ann. of Phys. {\bf 249}, 222--235 (1996).


\bibitem{Henkel:Godr02} 
C. Godr\`eche and J.-M. Luck, 
Nonequilibrium critical dynamics of ferromagnetic spin systems, 
J. Phys. Cond. Matt. {\bf 14}, 1589--1599 (2002); {\tt cond-mat/0109212}.

\bibitem{Henkel:Henk94} 
M. Henkel, 
Sch\"odinger invariance and strongly anisotropic critical systems
J. Stat. Phys. {\bf 75}, 1023--1061 (1994); {\tt hep-th/9310081}. 

\bibitem{Henkel:Henk99}
M. Henkel,
{\it Conformal Invariance and Critical Phenomena},
Heidelberg, Springer, 1999.

\bibitem{Henkel:Henk01b} 
M. Henkel, M. Pleimling, C. Godr\`eche and J.-M. Luck,
Ageing, phase ordering and conformal invariance, 
Phys. Rev. Lett. {\bf 87}, 265701 (2001); {\tt hep-th/0107122}.

\bibitem{Henkel:Henk02} 
M. Henkel, 
Phenomenology of local scale invariance: from conformal invariance to 
dynamical scaling, 
Nucl. Phys. {\bf B641}, 405--486 (2002); {\tt hep-th/0205256}.
 
\bibitem{Henkel:Henk02a} 
M. Henkel, M. Paessens and M. Pleimling, 
Scaling of the magnetic linear response in phase-ordering kinetics, 
Europhys. Lett. {\bf 62}, 664--670 (2003); {\tt cond-mat/0211583}.

\bibitem{Henkel:Henk02b} 
M. Henkel and M. Pleimling, 
Local scale invariance as dynamical space-time symmetry in phase-ordering 
kinetics, Phys. Rev. {\bf E68}, 065101(R) (2003); {\tt cond-mat/0302482}. 

\bibitem{Henkel:Henk03} 
M. Henkel and J. Unterberger, 
Schr\"odinger invariance and space-time symmetries, 
Nucl. Phys. {\bf B660}, 407--435 (2003); {\tt hep-th/0302187}. 

\bibitem{Henkel:Maze03}
G.F. Mazenko, 
Response functions in phase-ordering kinetics, Phys. Rev. {\bf E69}, 
016114 (2004); {\tt cond-mat/0308169}. 

\bibitem{Henkel:Nied72} 
U. Niederer, 
The maximal kinematical invariance group of the free Schr\"odinger equation,
Helv. Phys. Acta {\bf 45}, 802--810 (1972).

\bibitem{Henkel:Niki03} 
A.G. Nikitin and R.O. Popovych, 
Group classification of nonlinear Schr\"odinger equations, 
Ukr. Math. J. {\bf 53}, 1255--1265 (2001); {\tt math-ph/0301009}.

\bibitem{Henkel:ORai01} 
L. O'Raifeartaigh and V.V. Sreedhar, 
The maximal kinematical invariance group of fluid dynamics and 
explosion-implosion duality, 
Ann. of Phys. {\bf 293}, 215--227 (2001); {\tt hep-th/0007199}.

\bibitem{Henkel:Pico02} 
A. Picone and M. Henkel, 
Response of non-equilibrium systems with long-range initial correlations, 
J. Phys. {\bf A35}, 5575--5590 (2002); {\tt cond-mat/0203411}.

\bibitem{Henkel:Pico03} 
A. Picone and M. Henkel, Local scale-invariance and ageing in noisy systems;
{\tt cond-mat/0402196}.

\bibitem{Henkel:Sast03}
F. Sastre, I. Dornic and H. Chat\'e, 
Bona fide thermodynamic temperature in nonequilibrium kinetic Ising models,
Phys. Rev. Lett. {\bf 91}, 267205 (2003); {\tt cond-mat/0308178}. 

\bibitem{Henkel:Sche03} 
G. Schehr and P. Le Doussal, 
Exact multilocal renormalization on the effective action: application to 
the random sine-Gordon model statics and non-equilibrium dynamics, Phys. Rev.
{\bf E68}, 046101 (2003); {\tt cond-mat/0304486}. 

\bibitem{Henkel:Yeun96} 
C. Yeung, M. Rao and R.C. Desai, 
Bounds on the decay of the auto-correlation in phase-ordering dynamics, 
Phys. Rev. {\bf E53}, 3073--3077 (1996); {\tt cond-mat/9409108}.

\end{thebibliography}
\end{document}